\documentclass[aps,prc,twocolumn,floatfix,superscriptaddress]{revtex4}
\usepackage{epsfig}
\usepackage{amsmath}
\usepackage{color}

\usepackage{graphicx}

\begin{document}

\title{Directed flow of photons in Cu+Au collisions at RHIC}
\author{Pingal Dasgupta}
\email{pingaldg@vecc.gov.in}
\author{Rupa Chatterjee}
\email{rupa@vecc.gov.in}
\author{Dinesh K. Srivastava}
\email{dinesh@vecc.gov.in}
\affiliation{Variable Energy Cyclotron Centre, HBNI, 1/AF, Bidhan Nagar, Kolkata-700064, India}

\begin{abstract}
Event-by-event fluctuations in the positions of nucleons in two colliding identical nuclei can lead to non-uniform initial energy density distribution on the transverse plane. In addition to initial state fluctuations, the difference in the number of participating nucleons in collision of two non-identical nuclei can also 
result in significant anisotropy in the initial geometry and energy density distributions. Thus, Cu+Au collisions are expected to provide interesting new aspects in the understanding of anisotropic flow in heavy ion collisions.
We calculate directed flow co-efficient $v_1$ of thermal photons using a hydrodynamic model with fluctuating initial conditions at 200A GeV  Cu+Au collisions at RHIC and compare it with the elliptic and triangular flow parameters  obtained at same initial conditions. The photon $v_1$ as a function of transverse momentum  is found to be non-zero and significantly large. However, it shows a different nature compared to the elliptic and triangular flow parameters. The $v_1$ is found to be completely dominated by QGP radiation in the region $1 < p_T < 6$ GeV and  contribution from the hadronic phase to photon $v_1$ is only  marginal. At  $p_T < 2$ GeV, it is negative and it decreases further with smaller values of $p_T$.  However, at $p_T > 2 $ GeV, $v_1$ is  positive and rises slowly with $p_T$. In addition, the photon $v_1$ is found to be more sensitive to the initial formation time of the plasma  compared to the elliptic and triangular flow parameters.   We suggest that a simultaneous measurements of photon $v_n$ co-efficients, (n=1, 2, 3) can provide valuable information about the initial state produced in heavy ion collisions as well as help us understanding the direct photon puzzle.
\end{abstract}

\maketitle

\section{Introduction} 
Anisotropic flow of identified particles produced in ultra-relativistic heavy ion collisions is considered to be one of the most powerful signatures of the formation of hot and dense strongly interacting Quark Gluon Plasma (QGP) matter in those collisions. Relativistic hydrodynamic model with smooth initial density distribution successfully explained the particle spectra and elliptic flow of hadrons at RHIC initially~\cite{uli, hydro1}. However, soon it was realised that event-by-event hydrodynamic model with fluctuating initial conditions (IC) provided a better description of the experimental data. Thus it is now  considered to be more realistic than a smooth initial density distribution to explain the bulk properties of the medium produced in heavy ion collisions~\cite{hannu,pt,scott,hannah, sorenson,nex}. In addition, event-by-event fluctuating IC are found to explain the significantly large  elliptic flow produced in most central Cu+Cu collisions at RHIC and non-zero triangular flow of hadrons, both of which remained  unexplained earlier using smooth initial density distribution~\cite{cds}. 

Although the relativistic hydrodynamic model successfully explained the hadronic observables, the same model can not explain the photon spectra and anisotropic flow simultaneously. Experimental data for both the elliptic and triangular flow co-efficients from Au+Au collisions at RHIC and from Pb+Pb collisions at the LHC are found to be significantly larger compared to the result from hydrodynamical model calculation~\cite{phenix_phot,phenix_v2,alice_phot,alice_v2,chre3,chre4,cds}. This is known as {\it direct photon puzzle}. In recent  studies we have shown that calculation of photon anisotropic flow in collisions of deformed nuclei (U+U)~\cite{uu} as well as from smaller systems (Cu+Cu)~\cite{cds} could be valuable to understand this puzzle. 
Event-by-event fluctuating initial conditions produce lumpy initial state with non-uniform distribution of energy density in Au+Au and Pb+Pb collisions at RHIC and LHC energies. 
However, collision of non-identical Cu and Au nuclei are especially interesting as these may lead  to collision geometries which are asymmetric on the transverse plane even with smooth initial energy density distributions (see Fig. 1). 
It is quite well known that the odd flow co-efficients ($v_3$, $v_5$.. etc)  originate  due to the initial state fluctuations. On the other hand, the even flow co-efficients, ($v_2$ and $v_4$)  which are already present for smooth IC, get enhanced due to the fluctuations in the initial density distribution. We have not come across  many studies on the directed flow coefficient $v_1$, which originates due to collective side ward motion where the momentum distribution is shifted towards one of the sides on the transverse plane~\cite{star_v1even,v1_1even,alice_v1even}. Some earlier studies have shown that $v_1$ or directed flow originates only from collisions at the lower beam energies due to the side ward motion of the spectator nucleons~\cite{v1_odd1,v1_odd2}.   

Directed flow is expected to form at very early stages of heavy ion collisions and thus is sensitive to the initial pressure gradients of the evolving nuclear matter compared to the other higher order flow co-efficients~\cite{y.b.evanov, grassi hama kodama}. Typical time scale of $v_1$ is the overlapping time between the two colliding nuclei and this time scale  is expected to decrease with increasing beam energy.
Thermal emission of photons is highly sensitive to the initial temperature and formation time of QGP and thus photon $v_1$ could be a potential probe to study the initial state produced in heavy ion collisions. In addition, the asymmetric overlapping zone produced in collisions of Cu and Au nuclei are also expected to provide non-trivial information about  anisotropic flow produced in heavy ion collisions~\cite{bozek}. A recent study on photon $v_1$ from 200A GeV Au+Au collisions has shown that the largest $v_1$ signal comes from the region near the phase transition, at temperatures $T \sim 150 -− 200$ MeV~\cite{photon_v1_shen}.

We calculate $p_T$ dependent directed flow co-efficient $v_1$ of thermal photons from 200A GeV Cu+Au collisions at RHIC. Results are obtained at mid-rapidity and for 20--30\% centrality bin.  The elliptic and triangular flow parameters are also calculated using same initial conditions and results are compared with photon $v_1$. The individual contributions of the QGP and hadronic matter phases to total $v_n$ (n=1, 2, 3) are studied in detail. In addition, the sensitivity of the results to initial formation time of the plasma has been  examined. We argue that experimental determination of photon $v_n$ from Cu+Au collisions can be crucial in understanding the direct photon puzzle and also the initial state produced in relativistic heavy ion collisions.

\section{formalism}
\begin{figure}
\centerline{\includegraphics*[width=9.0 cm,clip=true]{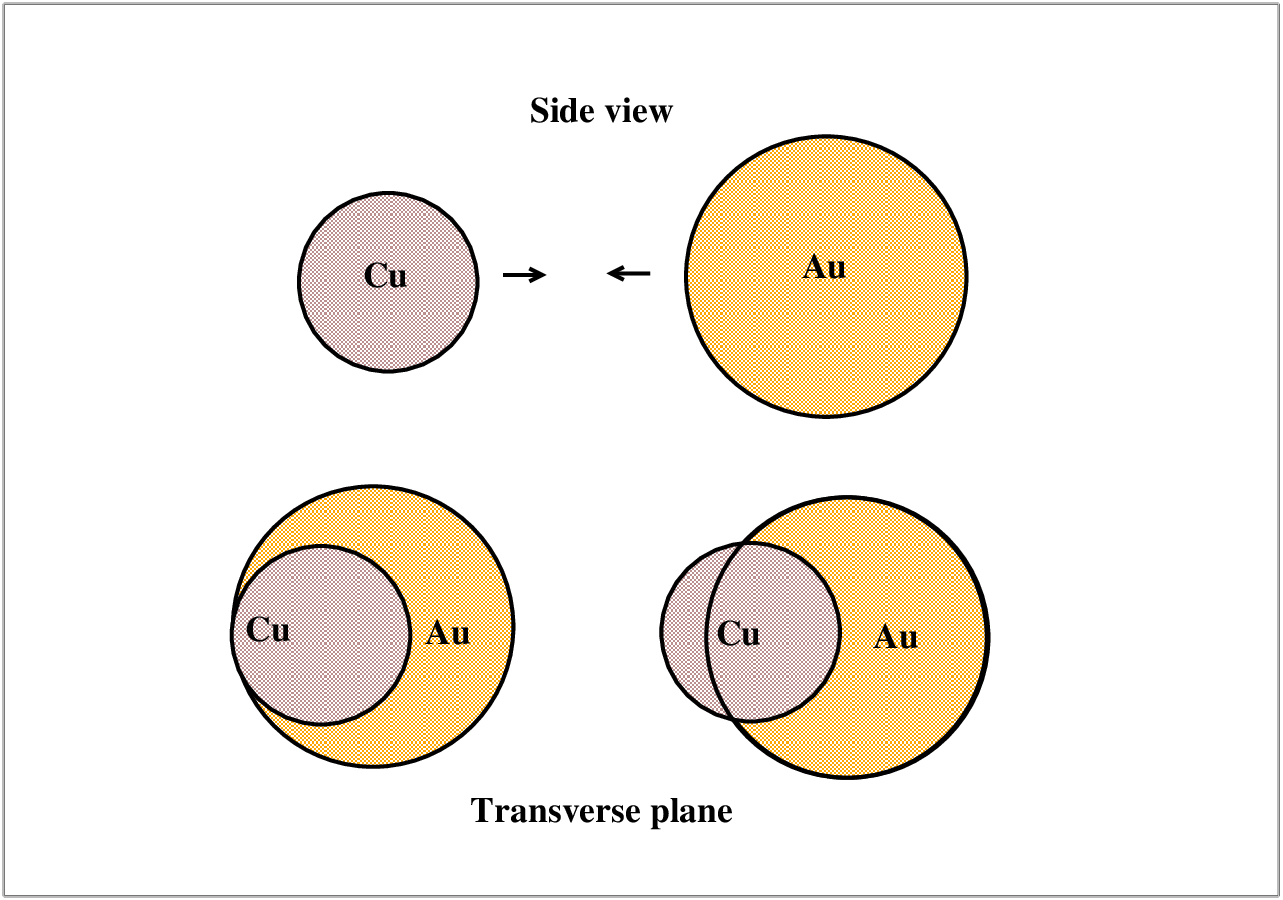}}
\caption{(Color online) Schematic of Cu and Au collisions for 20--30\% centrality class. }
\label{f1}
\end{figure}

We consider a (2+1) dimensional longitudinally boost invariant event-by-event ideal relativistic hydrodynamic framework~\cite{hannu} to model the evolution of the hot and dense QGP matter produced in the collision of Cu and Au nuclei at 200A GeV at RHIC. The photon $v_1$ is calculated considering a sufficiently large number of events from a centrality window of 20--30\% and Fig.\ref{f1} shows the (schematic) two different configurations of Cu+Au collisions in this centrality bin. 
We have checked that a 2+1 dimensional hydrodynamical model description is good enough to capture the effects of initial-state fluctuations with asymmetric participating nuclei from Cu and Au at mid-rapidity as the result are found to be similar to a 3+1 dimensional model calculation.  We consider a Monte Carlo Glauber initial condition to distribute the initial entropy densities.  The initial profile is taken as (proportional to) a linear combination of wounded nucleons (0.875) and binary collisions (0.125)~\cite{bozek}.  This framework has been used extensively earlier to calculate the production and anisotropic flow of photons at RHIC and LHC energies~\cite{chre1,chre2}.

A 2-dimensional Gaussian distribution function 

\begin{equation}
  s(x,y) = \frac{K}{2 \pi \sigma^2} \sum_{i=1} \exp \Big( -\frac{(x-x_i)^2+(y-y_i)^2}{2 \sigma^2} \Big)
\label{eq:eps}
\end{equation}
is used to distribute the initial energy/entropy densities. ($x_i, y_i$) denote the position of the nucleon on the transverse plane. The constant factor K in the equation above is fixed by reproducing the charged particle multiplicity ($dN_{\rm ch}/d\eta\approx$140 for 20--30\% collision centrality)~\cite{bozek} at RHIC for Cu+Au collisions. K is taken as 81.  The parameter $\sigma$ plays an important role as it decides the fluctuation size or the granularity in the initial energy density distribution. The initial formation time $\tau_0$ for Au+Au collisions was taken as  0.17 fm from EKRT mini-jet saturation model~\cite{ekrt} in our earlier studies. We take the same value of $\tau_0$ for Cu+Au collisions. However,  as mentioned earlier that photon $v_1$ from Cu+Au collisions can be quite sensitive to the initial state,  we  also calculate $v_1$ considering  a 3 times larger value of $\tau_0=$  0.51 fm in later part of our study. A default value of  $\sigma=$ 0.4 fm is considered~\cite{hannu,chre1,chre2}.

The production of photons from the plasma phase is estimated by using the complete leading order photon rates from~\cite{amy,nlo_thermal}. We use the parameterized rates from~\cite{trg} to calculate the photon production from the hot hadronic matter phase produced in Cu+Au collisions at 200A GeV. 
The constant freeze-out temperature ($T_F$) is taken as 160 MeV which reproduces the pion spectra at RHIC well and  the hydrodynamic equations are solved considering a  lattice based equation of state from~\cite{eos}.

The emission rates ($R=EdN/d^3pd^4x$) from QGP and hadronic matter phases are integrated over the space-time history to estimate the total thermal production and anisotropic flow  from sufficiently large number of events.

The anisotropic flow co-efficients $v_n$ (where n=1, 2 and 3)  are calculated by expanding the invariant particle distribution in transverse plane using Fourier decomposition:

\begin{equation}\label{eq: v2}
 \frac{dN}{d^2p_TdY} = \frac{1}{2\pi} \frac{dN}{ p_T dp_T dY}[1+ 2\, \sum_{n=1}^{\infty} v_n (p_T) \, \rm{cos} \, n (\phi - \psi_n^{PP})] \, .
\end{equation}
where the  participant plane angle $\psi_n^{\rm PP}$ in each event is calculated using the relation,
\begin{equation}
  \psi_{n}^{\text{PP}} = \frac{1}{n} \arctan 
  \frac{\int \mathrm{d}x \mathrm{d}y \; r^m \sin \left( n\phi \right) \varepsilon\left( x,y,\tau _{0}\right) } 
  { \int \mathrm{d}x \mathrm{d}y \; r^m \cos \left( n\phi \right) \varepsilon\left( x,y,\tau _{0}\right)}  + \pi/n \, .
\end{equation}
 $\epsilon (x,y,\tau _{0})$ is the energy density at time $\tau_0$ at (x,y) point on the transverse plane  and $r^m$ are weights to the energy density which are taken as $r^3$~\cite{bozek},$r^2$ and $r^2$~\cite{cds} for $\psi_1$,$\psi_2$  and $\psi_3$ respectively.

\section{results}
\begin{figure}
\centerline{\includegraphics*[width=8.0 cm]{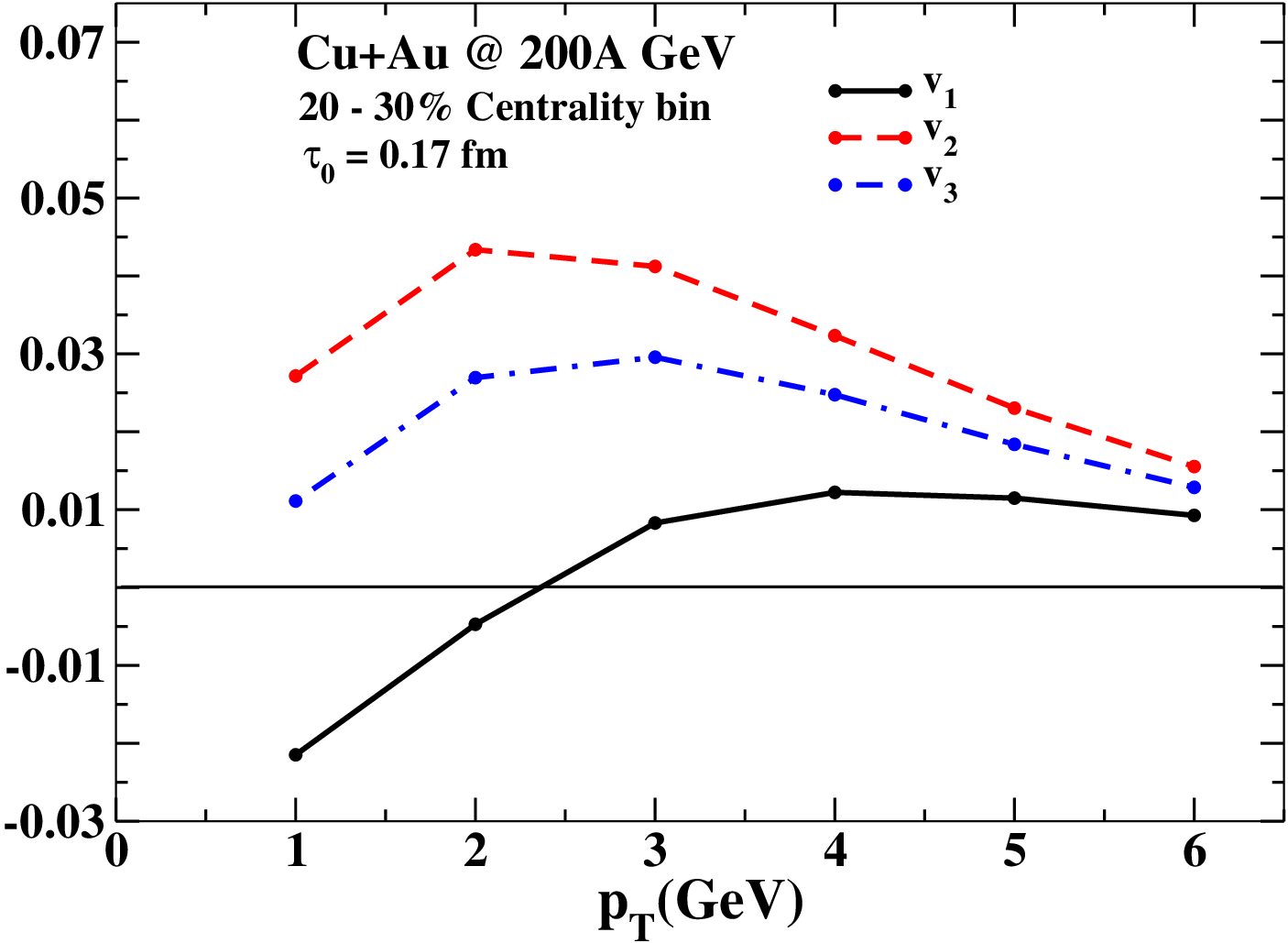}}
\caption{(Color online) Directed, elliptic, and triangular flow parameter of thermal photons as a function of $p_T$ from Cu+Au collisions at 200A GeV at RHIC and for centrality bin 20--30\%. }
\label{vn}
\end{figure}

The directed flow co-efficient $v_1$ of thermal photons as a function of $p_T$  for 20--30\% centrality bins at RHIC for Cu+Au collisions is shown in Fig.~\ref{vn}. The elliptic and triangular flow co-efficients are also shown in the same figure for a comparison. The results are obtained by taking the average over 400 fluctuating events from the centrality bin. 

The photon $v_2$ and $v_3$ from Cu+Au collisions is similar in nature to the flow parameters obtained earlier from Au+Au and Pb+Pb collisions at RHIC and LHC energies respectively. 
However, the directed flow co-efficient shows a nature different than the elliptic as well as the triangular flow parameters.  In the region $p_T \le$ 2.5 GeV,  $v_1$ is found to be negative whereas, for $p_T >$ 2.5 GeV the photon $v_1$ is positive and rises slowly with larger $p_T$. 
We have checked that the photon directed flow as a function of $p_T$ calculated from Au+Au collisions is much smaller compared to the result from Cu+Au collisions.

\begin{figure}
\centerline{\includegraphics*[width=8.0 cm]{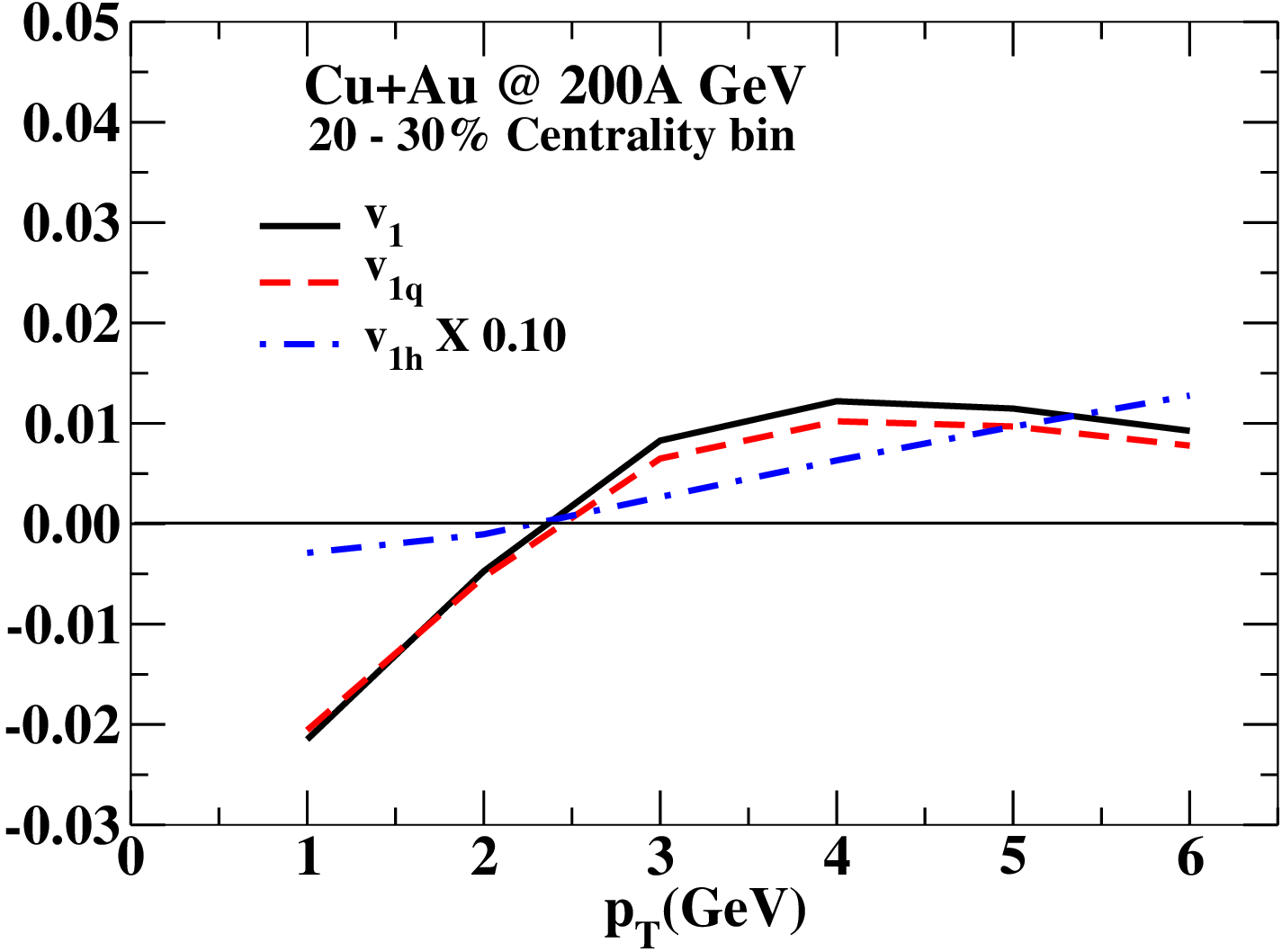}}
\caption{(Color online) Directed flow co-efficient $v_1$ as a function of $p_T$ of thermal photons along with separate contributions (to $v_1$) from the QGP and the hadronic matter phases. }
\label{v1qh}
\end{figure}

In order to understand the nature of $v_1$, individual contributions from the quark matter and hadronic matter phases to the total thermal photon $v_1$ are plotted in Fig.~\ref{v1qh}. 

One can see that the total photon $v_1$ is close to the  $v_1$ from  QGP (only) phase. The $v_1$ from hadronic phase  is much larger than the QGP $v_1$, however hadronic $v_1$ does not contribute significantly to total photon $v_1$. The photon directed flow  is calculated by taking appropriate weight factor (photon yield) from the plasma and hadronic matter phases and the QGP photons completely outshine the photons from hadronic phase in the region $p_T >$ 1 GeV. Thus, the photon $v_1$ is totally dominated by plasma radiation in the entire $p_T$ region shown in the figure.  This is quite different to the case of photon elliptic flow parameter where the hadronic matter contribution plays a significant role. For photon elliptic flow we have seen that although the total $v_2$  shows similar qualitative nature as the $v_2$ from quark matter, the sum $v_2$ is much larger than the contribution from quark matter. As a result photon $v_1$ is  expected to  probe the first few fm time period of the initial hot and dense state of matter better compared to the higher flow co-efficients $v_2$, $v_3$, etc.

\begin{figure}
\centerline{\includegraphics*[width=8.0 cm]{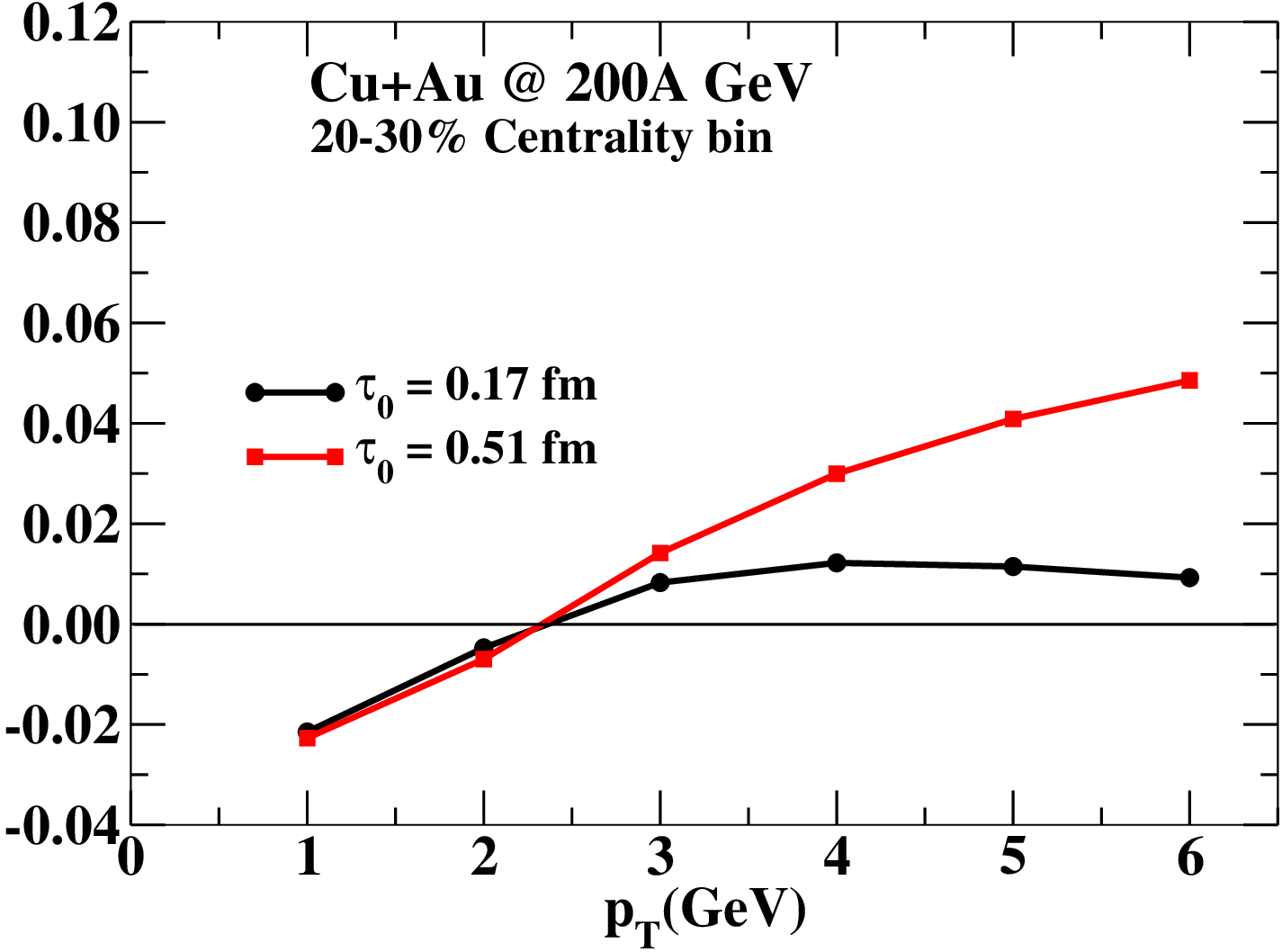}}
\caption{(Color online) Directed flow $v_1$ of thermal photons from Cu+Au collisions at RHIC for 20--30\% centrality bins for two different initial formation time $\tau_0$ as 0.17 fm and 0.51 fm.}
\label{v1_comp}
\end{figure}

The dependence of photon $v_1$ on initial formation time $\tau_0$ is shown in Fig.~\ref{v1_comp}. We compare the results obtained form $\tau_0$ value of 0.51 fm with the one at 0.17 fm keeping the entropy ($s\tau$) same for both the cases. In lower $p_T$ values where $v_1$ is negative, the two results are close to each other as this $p_T$ region is dominated by radiation from the relatively later stage of evolution. The difference in $v_1$ for the two $\tau_0$ values  is quite visible at larger $p_T$. Larger $\tau_0$ results in decrease of high $p_T$ photons (with smaller transverse flow velocity) from initial stage. This enhances the relative contribution of the flow boosted QGP photons emitted from relatively later stage of system evolution and as a result we see larger $v_1$ at high $p_T$ for larger $\tau_0$. 

\begin{figure}
	\centerline{\includegraphics*[width=8.0 cm]{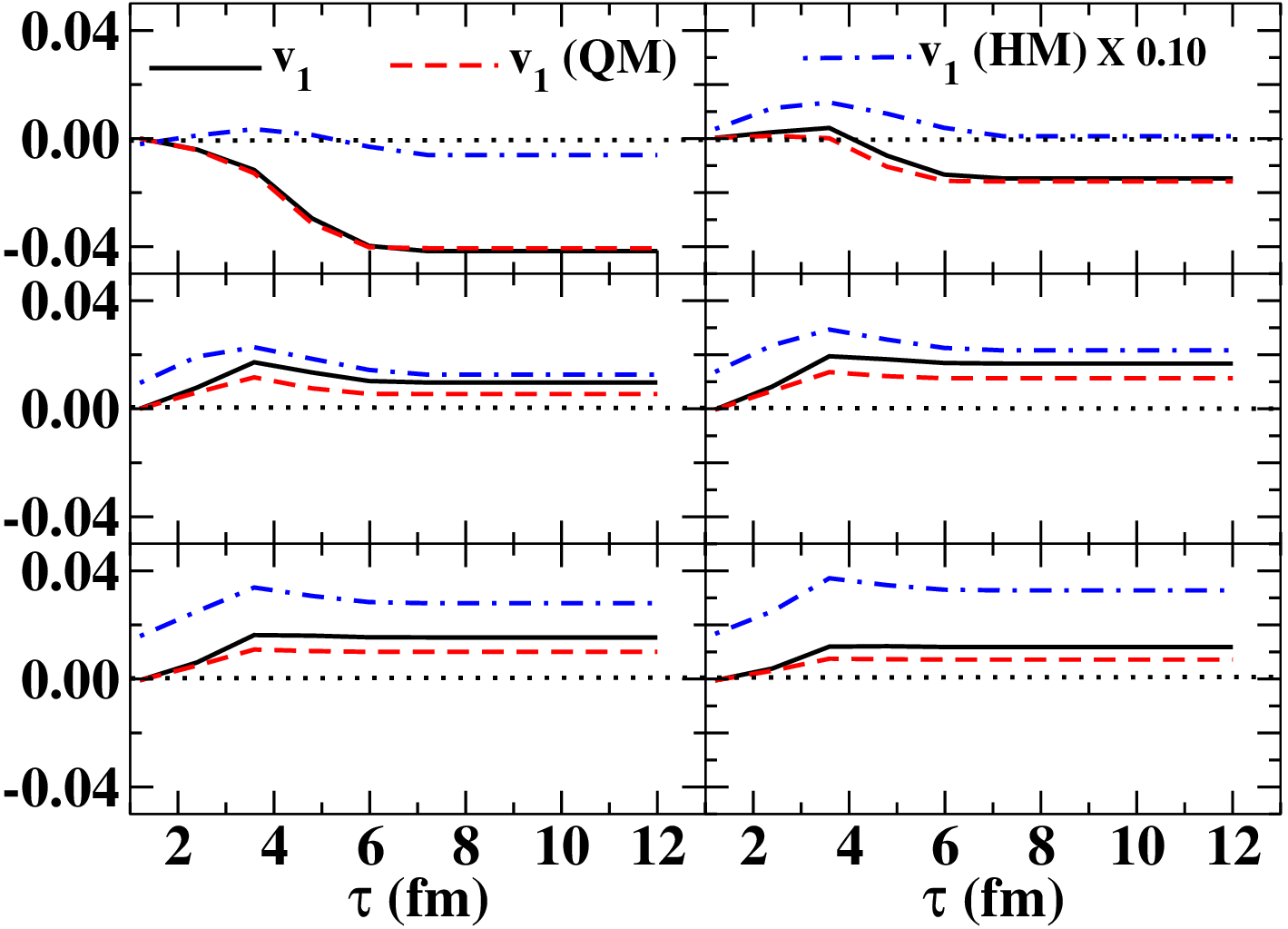}}
	\caption{(Color online) Time evolution of $v_1$ from QGP and hadronic matter phases.}
	\label{time_evolution}
\end{figure}

The gradual development of photon $v_1$  with time for  a single event is shown in Fig.~\ref{time_evolution}.  The photon $v_1(p_T)$ from this particular event is  found to be similar to the (400) event averaged $v_1 (p_T)$ shown in Fig.~\ref{vn}. The QGP and hadronic matter contributions are  also shown  separately  in the same figure.

 The  development of negative photon $v_1$ at smaller $p_T$ and positive photon $v_1$ at larger $p_T$ is result of the conservation of net transverse momentum of the fluid. The high $p_T$ photons are emitted along the direction of the steepest gradient of energy density whereas, the low $p_T$ photons, which counterbalance the net transverse momentum, are emitted in the opposite direction. In the beginning, the production of photons from the centre of the fireball is very large but flow developed by then is small. As a result, QGP contributions for all $p_T$ values are found to be small. However, at later times flow starts to develop slowly at the centre and the QGP $v_1$ is found to increase. The hadronic matter contribution is much larger compared to the QGP contribution and it shows interesting nature with time.  During the beginning of the expansion, the flow starts to develop from the boundary of the  fireball which is mostly in the hadronic states. As a result, hadronic photon $v_1$ starts to rise very rapidly in the positive direction for all $p_T$ values. However, the positive directed flow of hadronic photons with low momentum subsequently dies out and starts to grow in the negative direction which ultimately produces rather small negative hadronic photon $v_1$. At high $p_T$, hadronic photon $v_1$ is large but it is overshadowed by the QGP contributions.  Thus, we see a very small influence of hadronic contribution in the total photon $v_1$. 

\begin{figure}
\centerline{\includegraphics*[width=8.0 cm]{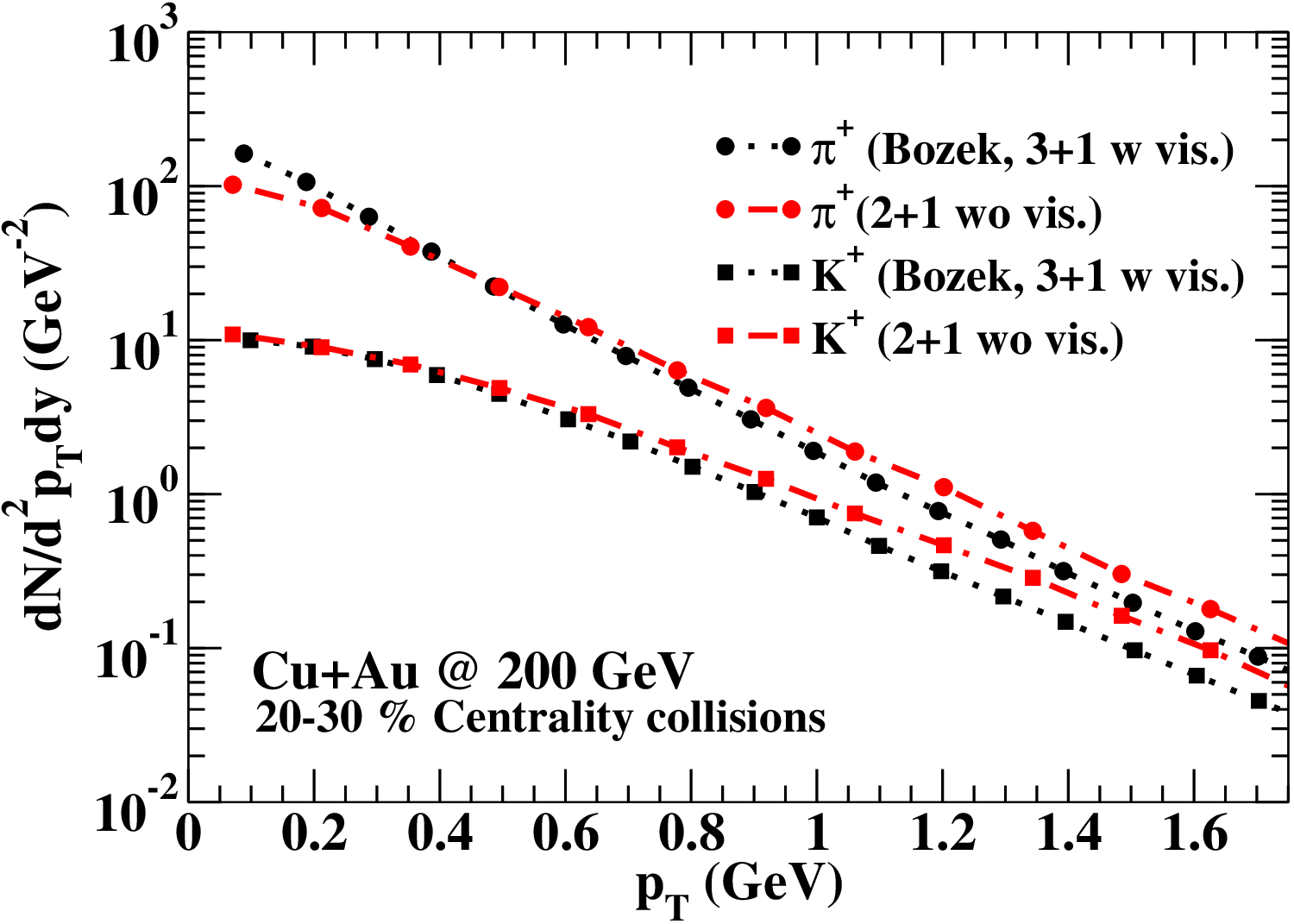}}
\caption{(Color online) $p_T$ spectra of pion and keons from Cu+Au collisions at RHIC.}
\label{hadronic}
\end{figure}

The predicted $p_T$ spectra of positive pions and kaons from 200A GeV Cu+Au collisions at RHIC and for 20--30\% centrality bin are shown in Fig.~\ref{hadronic}. We can not compare these hadronic spectra with experimental data due to unavailability of that at present. We have checked that our hadronic spectra from (2+1) dimensional hydrodynamic model calculation are close to the spectra obtained using a (3+1) dimensional viscous hydrodynamical model calculation in one recent study~\cite{bozek}. However, it is to be noted that an experimental comparison would be quite valuable.

\vspace{1.05cm}
\section{Summary and Conclusion}
We calculate the directed flow of thermal photons as a function of $p_T$ in Cu+Au collisions at 200A GeV at RHIC using an event-by-event hydrodynamic model with fluctuating initial conditions. The origin of $v_1$ is due to two reasons, one is the deformed overlapping geometry and the second one is fluctuations in the initial density distributions on the transverse plane.  The $p_T$ dependent directed flow of thermal photons at mid rapidity is found to be non-zero and shows a nature different than the elliptic and triangular flow parameter of thermal photons. The $v_1$ is found to be negative in the lower $p_T$ ( $< 2.5$ GeV) region and it  becomes positive for larger $p_T$ values. The photon $v_1$ is found to be entirely dominated by QGP radiation in the region $1<p_T<6$ GeV. In addition, directed flow builds up quite early, during the first few fm time period of the system evolution and thus is much more sensitive to the initial state compared to the higher order flow co-efficients. The time evolution of the flow-coefficient $v_1$ is explains the non trivial $p_T$ dependent  nature of it very well. 

We conclude that photon $v_1$ from Cu+Au collisions can be quite useful to understand the initial state produced in relativistic heavy ion collision and also the direct photon puzzle.

\begin{acknowledgments} 
We acknowledge the computer facility of Drona, Prafulla, and Physics clusters of VECC. DKS gratefully acknowledges the grant of Raja Ramanna Fellowship by the Department of Atomic Energy, India.
\end{acknowledgments}

\end{document}